\documentstyle[prb,aps,eqsecnum,preprint,tighten,floats,psfig]{revtex}

\begin{document}
\draft

\preprint{\vbox{\hfill UMN-D-01-8}  \\
          \vbox{\vskip0.3in}
          }

\title{Solution of the one-dimensional Dirac equation
with a linear scalar potential}

\author{John R. Hiller}
\address{Department of Physics,
University of Minnesota-Duluth, Duluth, Minnesota 55812}

\date{\today}

\maketitle

\begin{abstract}
We solve the Dirac equation in one space dimension for the
case of a linear, Lorentz-scalar potential.  This extends
earlier work of Bhalerao and Ram 
[Am.\ J. Phys.\ {\bf 69} (7), 817-818 (2001)]
by eliminating unnecessary constraints.  The spectrum
is shown to match smoothly to the nonrelativistic
spectrum in a weak-coupling limit.
\end{abstract}
\pacs{3.65.Pm, 3.65.Ge}

\narrowtext

\section{Introduction}

The linear potential $V(x)=g|x|$ is a natural choice for
a confining potential in one space dimension.  The nonrelativistic
Schr\"odinger equation admits a nearly analytic solution for this
potential in terms of an Airy function and the zeros of this function
and its derivative.  The Dirac equation, on the other hand, appears
to be problematic for this potential.  If $V$ is introduced as the
time component of a Lorentz two-vector, no bound-state solutions
exist.\cite{Galic,CapriFerrari}  If it is introduced as a
Lorentz scalar, Bhalerao and Ram\cite{BhaleraoRam} find only a
very limited set of solutions, with no obvious correspondence to
the nonrelativistic solutions.  Such an outcome in the scalar
case is unexpected because the Klein paradox is not a problem;
positive and negative energy particles both see a confining potential.
A nonrelativistic limit for the positive-energy solutions should
reproduce the known nonrelativistic spectrum.

This inconsistency in the scalar case can be 
resolved.\cite{deCastro}  The solution
found by Bhalerao and Ram\cite{BhaleraoRam} turns out to be
over constrained.  Here we will construct a more general solution
and show that the nonrelativistic results are recovered in an
appropriate limit.

To see that the Dirac-equation solution should match on to the
nonrelativistic solution, consider the equation (with $\hbar=1=c$)
\begin{equation}
[\alpha p+\beta(m+g|x|)]\psi=E\psi
\end{equation}
in a representation where\cite{representation}
\begin{equation}
\alpha\rightarrow\tilde{\alpha}\equiv\sigma_y
=\left(\begin{array}{rr} 0 & -i \\ i & 0 \end{array}\right)\,,\;\;
\beta\rightarrow\tilde{\beta}\equiv\sigma_z
=\left(\begin{array}{rr} 1 & 0 \\ 0 & -1 \end{array}\right)\,.
\end{equation}
As usual, let 
$\psi=\tilde\psi
\equiv\left(\begin{array}{c} \tilde{u} \\ \tilde{v}\end{array}\right)$
and decompose the matrix equation into two coupled equations
\begin{eqnarray} \label{eq:TildeCoupledEqns}
-\tilde{v}'+(m+g|x|)\tilde{u}&=&E\tilde{u}\,, \nonumber \\
 \tilde{u}'-(m+g|x|)\tilde{v}&=&E\tilde{v}\,.
\end{eqnarray}
For energies $E=m+\tilde{\varepsilon}$ near the rest mass $m$,
with $\tilde{\varepsilon}\ll m$, and for weak coupling $g\ll m^2$,
the second equation yields $\tilde{v}\simeq\tilde{u}'/2m$.
Substitution into the first equation brings
$-\tilde{u}''/2m+g|x|\tilde{u}\simeq\tilde{\varepsilon}\tilde{u}$,
which is immediately recognized as the nonrelativistic Schr\"odinger
equation.

The solution to the Schr\"odinger equation is obtained by noting
that $\tilde{u}''=2m(g|x|-\tilde{\varepsilon})\tilde{u}$
is the differential equation for a shifted 
Airy function.\cite{AbramowitzStegun}  This yields the
normalizable solution 
$\tilde{u}(x)={\cal N}\mbox{Ai}\left((2mg)^{1/3}[|x|-\tilde{\varepsilon}/g]\right)$,
with ${\cal N}$ a normalization constant.  Continuity of $\tilde{u}$
and $\tilde{u}'$ at $x=0$ requires that either 
$\mbox{Ai}'\left(-(2mg)^{1/3}\tilde{\varepsilon}/g\right)=0$
[for even solutions] or
$\mbox{Ai}\left(-(2mg)^{1/3}\tilde{\varepsilon}/g\right)=0$
[for odd solutions].
Let $-\rho_n$ and $-\rho'_n$ denote the nth zeros of
Ai and $\mbox{Ai}'$, respectively.  Then the nonrelativistic
eigenenergies are $\tilde{\varepsilon}'_n=\rho'_n(g^2/2m)^{1/3}$
for even solutions and $\tilde{\varepsilon}_n=\rho_n(g^2/2m)^{1/3}$
for odd.  The values of $\rho_n$ and $\rho'_n$ can be obtained 
from tables in Ref.~[\ref{ref:AbramowitzStegun}].  The first four
of each are as follows: $\rho_n=2.3381$, 4.0879, 5.5206, 6.7867
and $\rho'_n=1.0188$, 3.2482, 4.8201, 6.1633.

We would expect the Dirac equation to yield these same results in
the limit of weak coupling.  To see that negative energy solutions
do not cause any difficulties, we can use the methods of 
Coutinho, Nogami, and Toyama\cite{Coutinho} to prove the
following extension of their theorem B: For a scalar potential
that is everywhere nonnegative, the positive energy solutions
have energy $E\geq m$ and the negative energy solutions have
$E\leq -m$.  The proof depends on the freedom to pick as real the
solutions $\tilde{u}$ and $\tilde{v}$ to the coupled equations
(\ref{eq:TildeCoupledEqns}).  Inner products of $\tilde{u}$ and 
$\tilde{v}$ with the terms of these equations yield
\begin{eqnarray} 
\int\tilde{v}\tilde{u}'dx
   +\int\tilde{u}(m+V)\tilde{u}dx&=&E\int\tilde{u}^2 dx \,, \nonumber \\
\int\tilde{v}\tilde{u}'dx
   -\int\tilde{v}(m+V)\tilde{v}dx &=&E\int\tilde{v}^2 dx \,,
\end{eqnarray}
where $g|x|$ has been replaced by a generic scalar potential $V$
and an integration by parts has been performed in the first
term of the first equation.  For positive $E$ and $V$, the
second equation implies that $\int\tilde{v}\tilde{u}'dx\geq 0$
and then the first equation yields $E\geq m$.  Analogous steps
for negative $E$ yield $\int\tilde{v}\tilde{u}'dx\leq 0$
and $E\leq -m$.  Thus the two parts of the spectrum are
completely separate, and we are allowed to focus on the 
positive-energy solutions only.

\section{Solution of the Dirac Equation}

To solve the Dirac equation directly, we use the same
representation as Bhalerao and Ram,\cite{BhaleraoRam}
that is
\begin{equation}
\alpha\equiv\sigma_y
=\left(\begin{array}{rr} 0 & -i \\ i & 0 \end{array}\right)\,,\;\;
\beta\equiv\sigma_x
=\left(\begin{array}{rr} 0 & 1 \\ 1 & 0 \end{array}\right)\,.
\end{equation}
For $\psi=\left(\begin{array}{c} u \\ v \end{array}\right)$
they obtain the coupled equations
\begin{eqnarray} \label{eq:CoupledEqns}
 u'+(m+g|x|)u&=&Ev\,, \nonumber \\
-v'+(m+g|x|)v&=&Eu\,.
\end{eqnarray}
These equations decouple in terms of their variable 
$\xi=\sqrt{g}(m/g+|x|)$, such that for $x>0$,
\begin{eqnarray}
\left(-\frac{d^2}{d\xi^2}+\xi^2\right)u=(E^2/g+1)u\,, \nonumber \\
\left(-\frac{d^2}{d\xi^2}+\xi^2\right)v=(E^2/g-1)v\,,
\end{eqnarray}
and for $x<0$,
\begin{eqnarray}
\left(-\frac{d^2}{d\xi^2}+\xi^2\right)u=(E^2/g-1)u\,, \nonumber \\
\left(-\frac{d^2}{d\xi^2}+\xi^2\right)v=(E^2/g+1)v\,.
\end{eqnarray}
Obviously these are harmonic-oscillator-type equations.  The
normalizable solutions can be constructed from the Hermite 
functions\cite{Lebedev} of order $\nu$ and $\nu+1$, where 
$E^2=2(\nu+1)g$, as\cite{BhaleraoRam}
\begin{eqnarray} \label{eq:DiracSoln}
u&=&\left\{\begin{array}{ll} 
                  Ce^{-\xi^2/2}H_{\nu+1}(\xi)\,, & x>0 \\
                  C'\frac{E}{\sqrt{g}}e^{-\xi^2/2}H_\nu(\xi)\,, & x<0\,,
           \end{array} \right. \nonumber \\
v&=&\left\{\begin{array}{ll} 
                  C\frac{E}{\sqrt{g}}e^{-\xi^2/2}H_\nu(\xi)\,, & x>0 \\
                  C'e^{-\xi^2/2}H_{\nu+1}(\xi)\,, & x<0\,.
           \end{array} \right. 
\end{eqnarray}
However, because $\xi$ is always positive, $\nu$ is not restricted
to being an integer.\cite{HalfOscillator}

Continuity at $x=0$ requires that 
$CH_{\nu+1}(\alpha)=C'EH_\nu(\alpha)/\sqrt{g}$
and $C'H_{\nu+1}(\alpha)=CEH_\nu(\alpha)/\sqrt{g}$,
where $\alpha\equiv m/\sqrt{g}$ is defined as in
Ref.~[\ref{ref:BhaleraoRam}].  These conditions yield
$C'=\pm C$ and
\begin{equation} \label{eq:EigenCondition}
H_{\nu+1}(\alpha)=\pm\frac{E}{\sqrt{g}}H_\nu(\alpha)\,,
\end{equation}
the latter being the eigenvalue condition.  This condition
has a rich set of solutions when free of the restriction
to integer $\nu$; there are infinitely many solutions for any 
positive value of $\alpha$.

The sign that appears in the eigenvalue condition
(\ref{eq:EigenCondition}) corresponds
to the parity of the solution.  The parity operator\cite{BjorkenDrell} is 
reflection in $x$ combined with multiplication by the Dirac matrix 
$\beta$.  Because $\xi$ is independent of the sign
of $x$, we find that 
\begin{equation}
\beta\left(\begin{array}{l} u(-x) \\ v(-x) \end{array}\right)
=\left(\begin{array}{l} v(-x) \\ u(-x) \end{array}\right)
=\pm\left(\begin{array}{l} u(x) \\ v(x) \end{array}\right)\,.
\end{equation}
The presence of such a symmetry is, of course, necessary for the
match to the nonrelativistic solution.

\section{Nonrelativistic limit}

To recover the nonrelativistic solution, we must take an
appropriate limit.  We write $E=m(1+\varepsilon)$ and consider
small $\varepsilon$ as well as small $g$.  The latter corresponds
to large $\alpha$ and large $\nu$.  In the limit of large $\nu$,\
we find from Ref.~[\ref{ref:AbramowitzStegun}] that $H_\nu$ has
the asymptotic form
\begin{equation}
H_\nu(\xi)\sim 2^\nu e^{\xi^2/2}\Gamma\left(\frac{\nu+1}{2}\right)
    \left(\frac{t_\nu}{z_\nu^2-1}\right)^{1/4}\mbox{Ai}(t_\nu)\,,
\end{equation}
where $z_\nu=\xi/\sqrt{2\nu+1}$ and, for $z_\nu\leq 1$,
\begin{equation}
t_\nu=-\left(\frac{3}{4}(2\nu+1)
      \left[\cos^{-1}z_\nu-z_\nu\sqrt{1-z_\nu^2}\right]\right)^{2/3}\,.
\end{equation}
This immediately looks promising because the desired Airy function is
present.  We next expand in powers of $\alpha^{-1}$ and $\varepsilon$, 
with $y_\pm\equiv\varepsilon-\sqrt{g}|x|/\alpha\pm 1/2\alpha^2$
and use of $\cos^{-1}(1-y)\simeq\sqrt{2y}+\sqrt{y^3/72}$, to obtain
\begin{eqnarray}
z_\nu&=&\frac{1+\sqrt{g}|x|/\alpha}{\sqrt{1+2\varepsilon+\varepsilon^2-1/\alpha^2}}
    \simeq 1-y_-\,, \;\;
z_{\nu+1} \simeq 1-y_+ \\
t_\nu & \simeq & -\left(\frac{2m^4}{g^2}\right)^{1/3}y_-\,, \;\;
t_{\nu+1} \simeq -\left(\frac{2m^4}{g^2}\right)^{1/3}y_+\,.
\end{eqnarray}
Note that $\sqrt{g}|x|/\alpha$ is of order $\varepsilon$ at the classical
turning point, which sets a natural scale for $x$,
and that the nonrelativistic correspondence implies
that $\varepsilon$ is of order $(g^2/m^4)^{1/3}\sim\alpha^{-4/3}$.
Thus the $\varepsilon^2$ terms can be dropped relative to $\alpha^{-2}$.

At lowest order, these expansions imply
\begin{equation}
e^{-\xi^2/2}H_\nu(\xi),\,e^{-\xi^2/2}H_{\nu+1}(\xi)\sim
   \mbox{Ai}\left((2m^4/g^2)^{1/3}[|x|/\alpha-\varepsilon]\right)\,.
\end{equation}
To compare with the original nonrelativistic reduction we must
connect the two representations of the Dirac matrices. They are
related by a unitary transformation
\begin{equation}
U=\frac{i}{\sqrt{2}}\left(\begin{array}{rr} 1 & 1 \\ -1 & 1 \end{array}\right)\,,
\end{equation}
such that 
\begin{equation}
\tilde{\psi}=\left(\begin{array}{cc} \tilde{u} \\ \tilde{v} \end{array}\right)=
U\psi=U\left(\begin{array}{cc} u \\ v \end{array}\right)\,.
\end{equation}
Therefore, we have $\tilde{u}=\frac{i}{\sqrt{2}}(u+v)$ and
$\tilde{v}=\frac{i}{\sqrt{2}}(v-u)$, with $u$ and $v$ given
by (\ref{eq:DiracSoln}).  Thus for large $\alpha$,
$\tilde{u}$ does indeed reduce to an Airy function with the 
correct argument, given $\varepsilon=\tilde\varepsilon/m$
and $\alpha=m/\sqrt{g}$.

At the next order we obtain, with the aid of Stirling's 
formula\cite{AbramowitzStegun} for $\Gamma\left(\frac{\nu+2}{2}\right)$,
\begin{equation}
H_\nu(\xi)\sim 2^{\nu-1/4} e^{\xi^2/2}\Gamma\left(\frac{\nu+1}{2}\right)
    \left(\frac{2m^4}{g^2}\right)^{1/12}
          (1+y_-/4) \mbox{Ai}\left(-(2m^4/g^2)^{1/3}y_-\right)
\end{equation}
and
\begin{equation}
H_{\nu+1}(\xi)\sim 2^{\nu+3/4} e^{\xi^2/2}\sqrt{\frac{\nu+1}{2}}
         \Gamma\left(\frac{\nu+1}{2}\right)
    \left(\frac{2m^4}{g^2}\right)^{1/12}
          (1+y_+/4) \mbox{Ai}\left(-(2m^4/g^2)^{1/3}y_+\right)\,.
\end{equation}
The combination that appears in the eigenvalue condition
\[ 0=H_{\nu+1}(\alpha)\mp \sqrt{2\nu+2}H_\nu(\alpha) \]
then reduces to 
\begin{eqnarray}
0\simeq && \sqrt{2\nu}2^{\nu-1/4} e^{\alpha^2/2}
             \Gamma\left(\frac{\nu+1}{2}\right)
               \left(\frac{2m^4}{g^2}\right)^{1/12} \\
&& \times  \left[(1+y_-/4)\mbox{Ai}\left(-(2m^4/g^2)^{1/3}y_+\right)
          \mp(1+y_+/4)\mbox{Ai}\left(-(2m^4/g^2)^{1/3}y_-\right)
          \right]_{x=0}\,.  \nonumber
\end{eqnarray}
First-order Taylor expansions of the Airy functions about
$-(2m^4/g^2)^{1/3}\varepsilon$ yield
\begin{eqnarray} 
0\simeq && \mbox{Ai}\left(-(2m^4/g^2)^{1/3}\varepsilon\right)
-(2\alpha)^{-2/3}\mbox{Ai}'\left(-(2m^4/g^2)^{1/3}\varepsilon\right) \\
&& \mp\left[
  \mbox{Ai}\left(-(2m^4/g^2)^{1/3}\varepsilon\right)
+(2\alpha)^{-2/3}\mbox{Ai}'\left(-(2m^4/g^2)^{1/3}\varepsilon\right)
      \right]\,, \nonumber
\end{eqnarray}
where terms of order higher than $\alpha^{-2/3}$ have been dropped.
For even parity (the upper sign) we have
$\mbox{Ai}'\left(-(2m^4/g^2)^{1/3}\varepsilon\right)\simeq 0$
and for odd parity,
$\mbox{Ai}\left(-(2m^4/g^2)^{1/3}\varepsilon\right)\simeq 0$,
which are the nonrelativistic eigenvalue conditions.
Therefore the eigenvalues will match in the limit of
large $\alpha$ and small $\varepsilon$.

\section{Results and conclusions}

\begin{figure}[htbp]
\begin{tabular}{cc}
\psfig{file=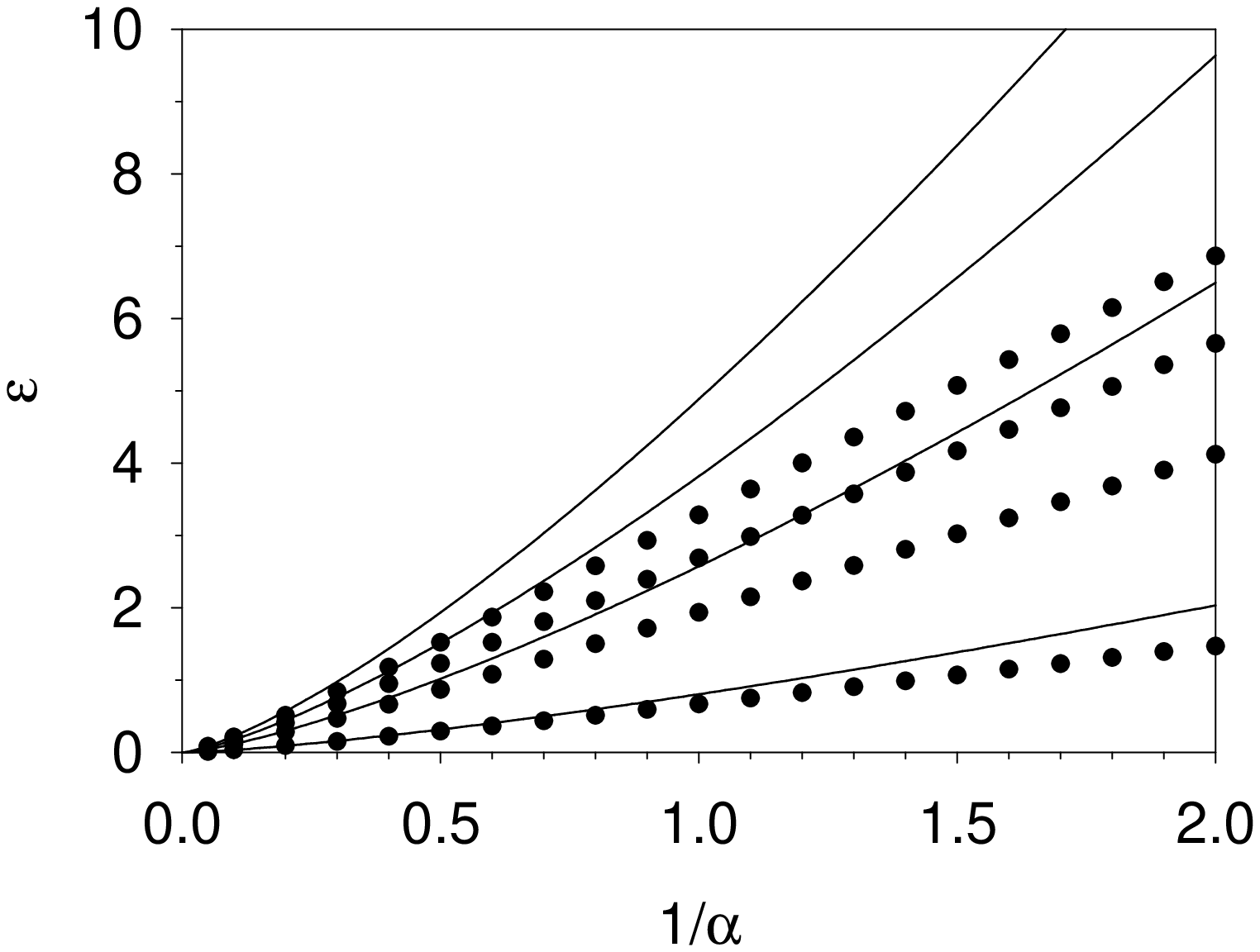,width=3.2in} &
\psfig{file=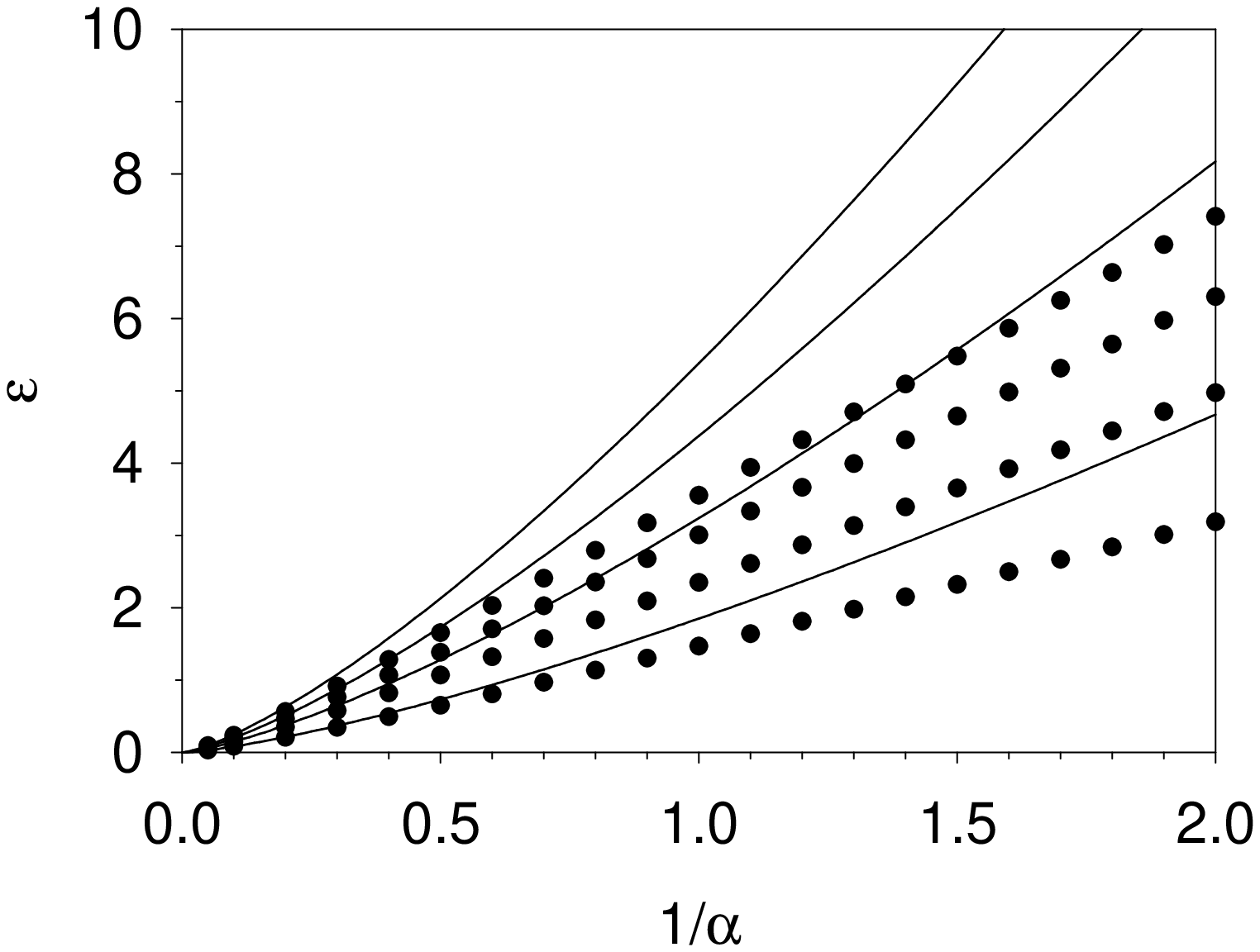,width=3.2in}  \\
(a) & (b)
\end{tabular}
\caption{\label{fig:eigenvalues} Lowest four
eigenvalues $\varepsilon\equiv E/m-1$ for the scalar potential as functions
of $1/\alpha\equiv \sqrt{g}/m$ for (a) even and (b) odd parity.
The solid lines are the nonrelativistic results,
and the points are positive-energy relativistic results.}
\end{figure}

Comparison of the relativistic and nonrelativistic eigenvalues
is made in Fig.~\ref{fig:eigenvalues}, where each is plotted
as a function of $1/\alpha$ for the four lowest levels
of each parity.  The nonrelativistic values were already
obtained above as explicit functions of $\alpha$ which are
simply plotted as lines in the figure.  The relativistic
values were computed numerically, with use of {\sc Mathematica}
to solve the eigenvalue condition (\ref{eq:EigenCondition}) at selected
values of $\alpha$.  The dimensionless $\varepsilon$ is related
to $\nu$ by $\varepsilon=\sqrt{2\nu+2}/\alpha-1$.  For $1/\alpha$
near 0, {\em i.e.}\ large $\alpha$, the relativistic and
nonrelativistic results are indistinguishable.  As $\alpha$
decreases, they separate smoothly.

From these results we see that the Dirac equation with a 
scalar linear potential is a well-defined problem in one dimension,
with a rich set of solutions and a smooth nonrelativistic limit.
As an exercise, one could extend this work to include calculation
of the relativistic wave functions and make direct comparisons with the
nonrelativistic Airy functions.  They will match in the 
large-$\alpha$ limit.  A second interesting exercise is a comparison
of the ultrarelativistic, strong-coupling limit of the spectrum to
the $\alpha^{-4/3}$ behavior of the nonrelativistic spectrum.  The
plots in Fig.~\ref{fig:eigenvalues} appear to imply a $\alpha^{-1}$
behavior for small $\alpha$.




\begin{references}
\bibitem{representation} The tildes are used to distinguish
this representation from the one used in Ref.~[\ref{ref:BhaleraoRam}].
\bibitem{HalfOscillator} The usual solution of odd-order 
Hermite polynomials for the half-harmonic oscillator, where
$V=\frac{1}{2}m\omega^2x^2$ for $x>0$ but is infinite for $x<0$,
is correct because the solutions to $H_\nu(0)=0$ 
are the odd integers, not because $\nu$ is first restricted to
being an integer by boundary conditions.
\end{references}
\end{document}